\documentclass{svproc}
\usepackage{url}

\usepackage{epsfig}
\usepackage{graphicx}
\usepackage{subfigure}
\usepackage{caption}
\usepackage{epstopdf}
\usepackage{multicol}
\usepackage{footmisc}
\usepackage{cite}

\begin{document}
\mainmatter              
\title{Time resolution and characteristic study of MWPC detectors with different Argon based gas}
\titlerunning{MWPC Characteristic}  
%
\author{Rajendra~Nath~Patra\inst{1} \and R.~N.~Singaraju\inst{1} \and
S.~Biswas\inst{2} \and T.~K.~Nayak\inst{1,3} \and Y.~P.~Viyogi\inst{1}}
\authorrunning{Rajendra Nath Patra et al.} 
%
\tocauthor{Rajendra Nath Patra, R. N. Singaraju,
	S. Biswas, T. K. Nayak, and Y. P. Viyogi}
\institute{Variable Energy Cyclotron Centre, HBNI, Kolkata-64, India,\\
\email{rajendrapatra07@gmail.com}
\and
Bose Institute, 93/1 APC Road, Kolkata-700009, India
\and
CERN, Geneva~23, Switzerland}
\maketitle              

\begin{abstract}
	A set of small scale prototypes of Multi-Wire Proportional Chambers (MWPCs) has been fabricated for the study of various characteristics. The detectors have been operated with $Ar/CO_{2}$ gas mixture having 70:30 and 90:10 ratio. Detector characteristic like efficiency, gain and time resolution have been studied using radioactive sources.
\end{abstract}
\section{Introduction}
The development of MWPC by Charpak~\cite{charpak1968} revolutionized the field of experimental nuclear and high energy physics. It found immense application in medical imaging also~\cite{lacymedical, chumedical}. In VECC, Kolkata we have developed many small scale prototype of MWPC detector for the characteristic study in terms of its efficiency, gain and time resolution with various gas mixtures.
\section{Experimental details and the test results}
The MWPC detectors contains an anode wire plane in between two conducting planes named as drift plane and the readout plane. The distance of the anode plane is 3 mm from both the drift and readout plane. Anode wire diameter is 20 $\mu$m and wire spacing is 2.8 mm. Details about the detector are given in~\cite{rajendra}. The detector is tested using Ar/CO$_{2}$ gas mixture having 70:30 and 90:10  ratio in flow mode at laboratory temperature and pressure.
\subsection{Efficiency and gain}
\begin{figure}[th!]
	\vspace{-3mm}
	\centering
	\includegraphics[width=2.6in]{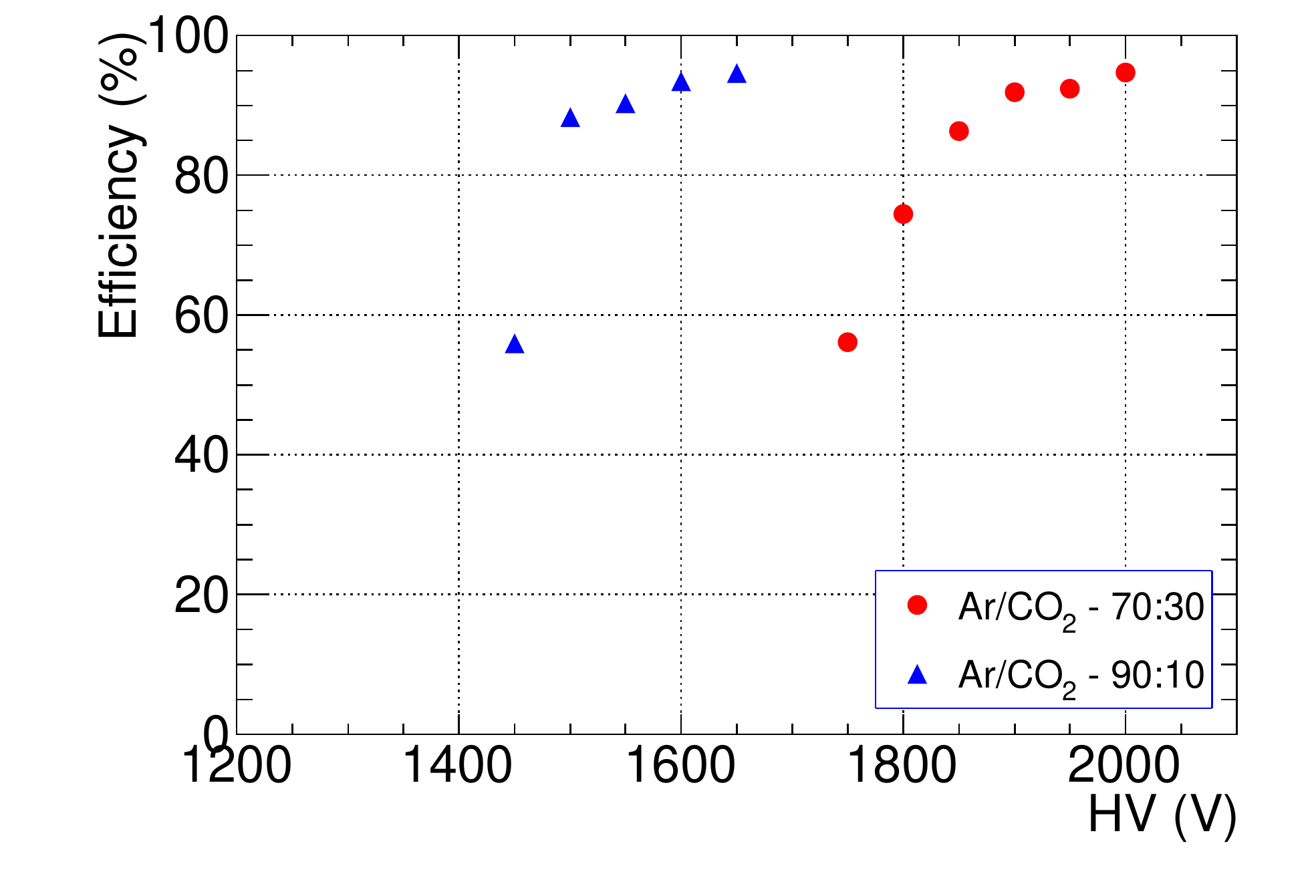}
	\vspace{-5mm}
	\caption{Efficiency as a function of HV in Ar/CO$_{2}$ 70:30 and 90:10 gas mixtures.}
	\label{efficiency}
	\vspace{-5mm}
\end{figure}

The efficiency of the detector is measured with $^{106}$Ru $\beta$-source using
a 3-fold trigger setup made up with three scintillator detectors - two crossed scintillators above and one scintillator below the MWPC under study. The efficiency as a function of applied high voltage (HV) is shown in Fig.~\ref{efficiency} for both gas mixtures. In the plateau region the efficiency is $\sim$94\% in both cases.

\begin{figure}[!th]
	\vspace{-5mm}
	\centering
	\includegraphics[width=2.35in]{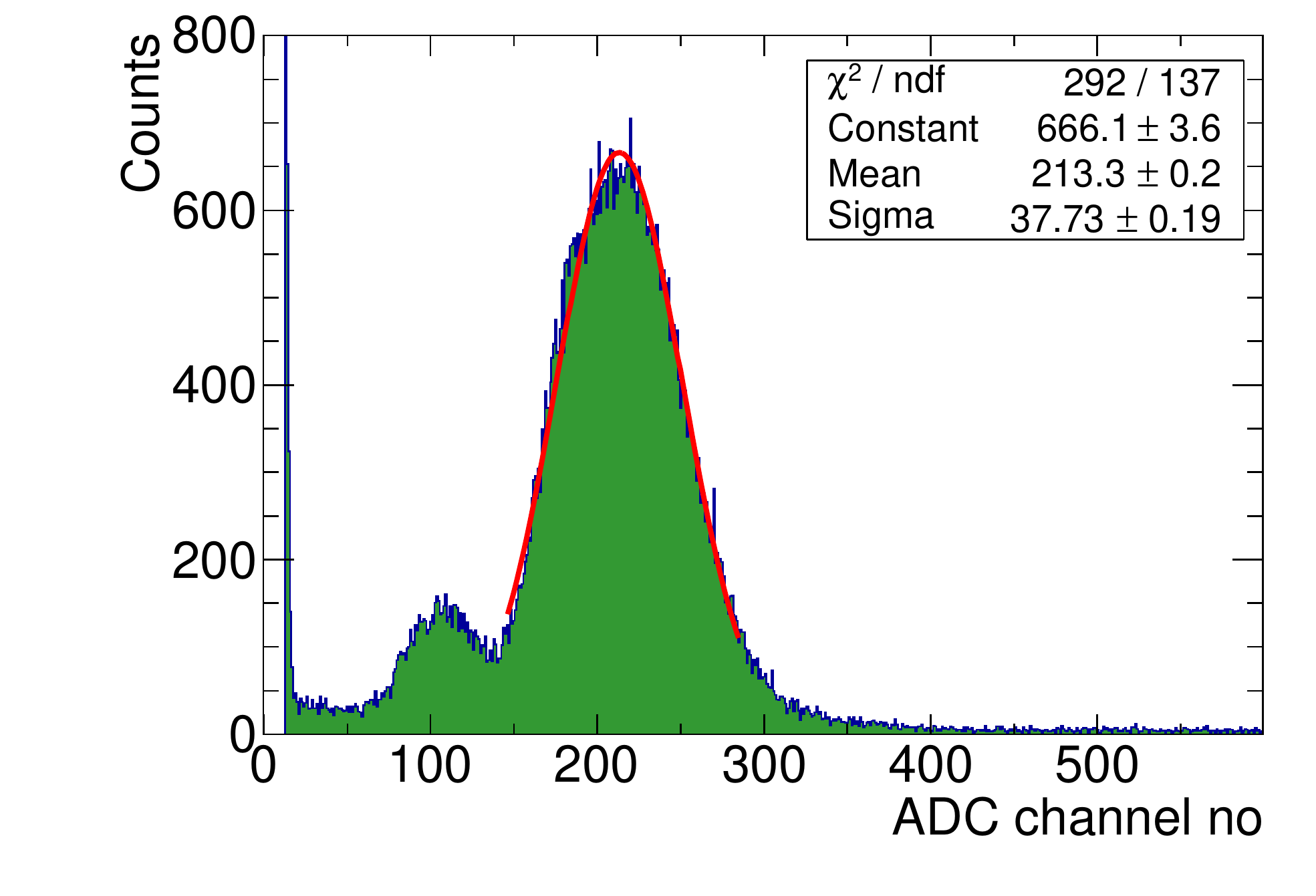}
	\includegraphics[width=2.35in]{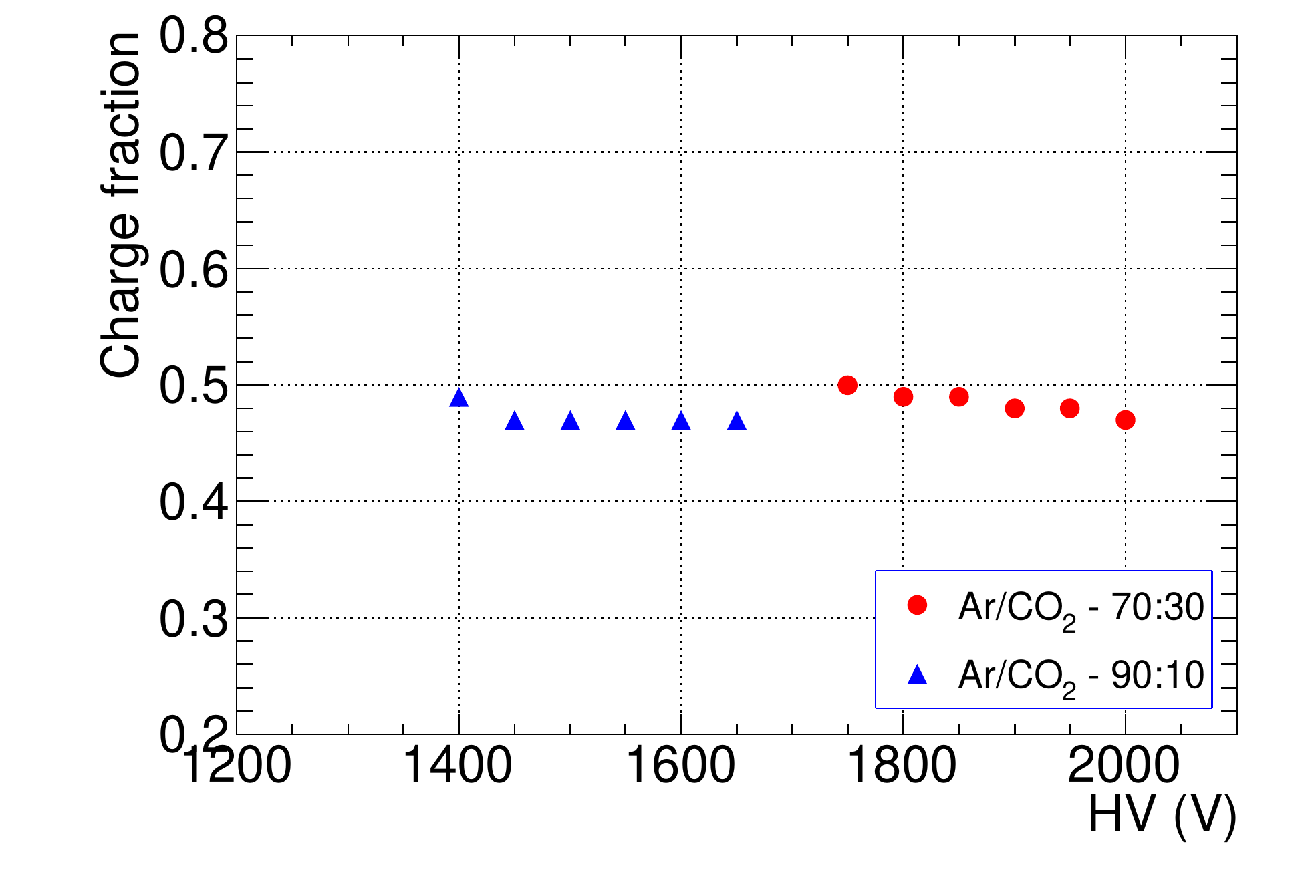}
	\vspace{-5mm}
	\caption{(a) $^{55}$Fe energy spectrum at 1900~V in Ar/CO$_{2}$ 70:30 gas, 
		(b) Charge fraction induced in the readout.}
	\label{fe55 chrg}
\vspace{-5mm}
\end{figure}

The detector is also tested using $^{55}$Fe 5.9~keV X-ray source to measure the gas gain. The anode energy spectrum of $^{55}$Fe at 1900 V in Ar/CO$_{2}$ (70:30) gas is given in Fig.~\ref{fe55 chrg}(a). The anode signal  induces an opposite polarity signal in the readout. The induced charge is shared by the readout and the drift plane equally. We measured the charge sharing fraction. This is shown in 
Fig.~\ref{fe55 chrg}(b).  The charge fraction is found to be $\sim$48\% and $\sim$47\% for the two cases. 

\begin{figure}[!th]
	\vspace{-5mm}
\centering
\includegraphics[width=2.35in]{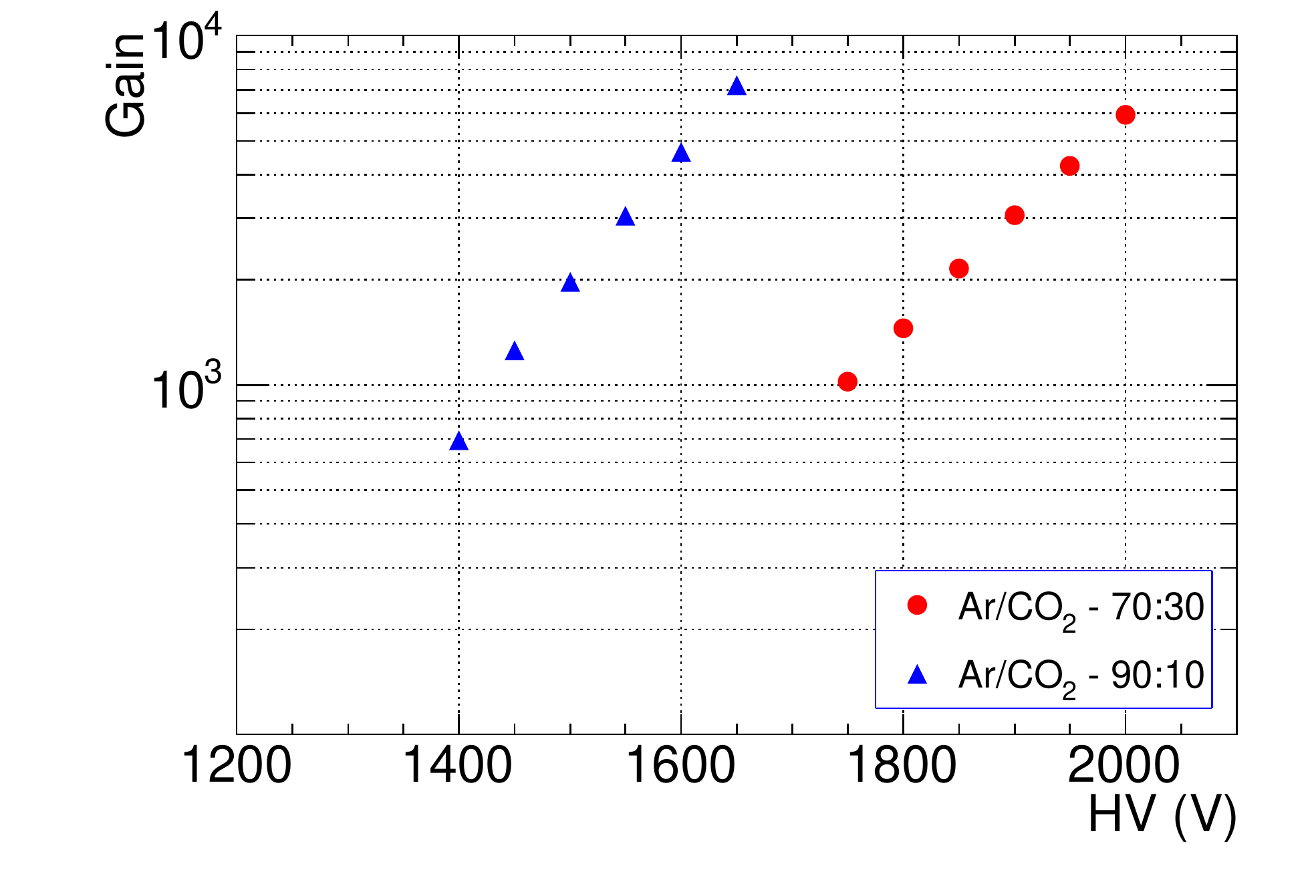}
\includegraphics[width=2.35in]{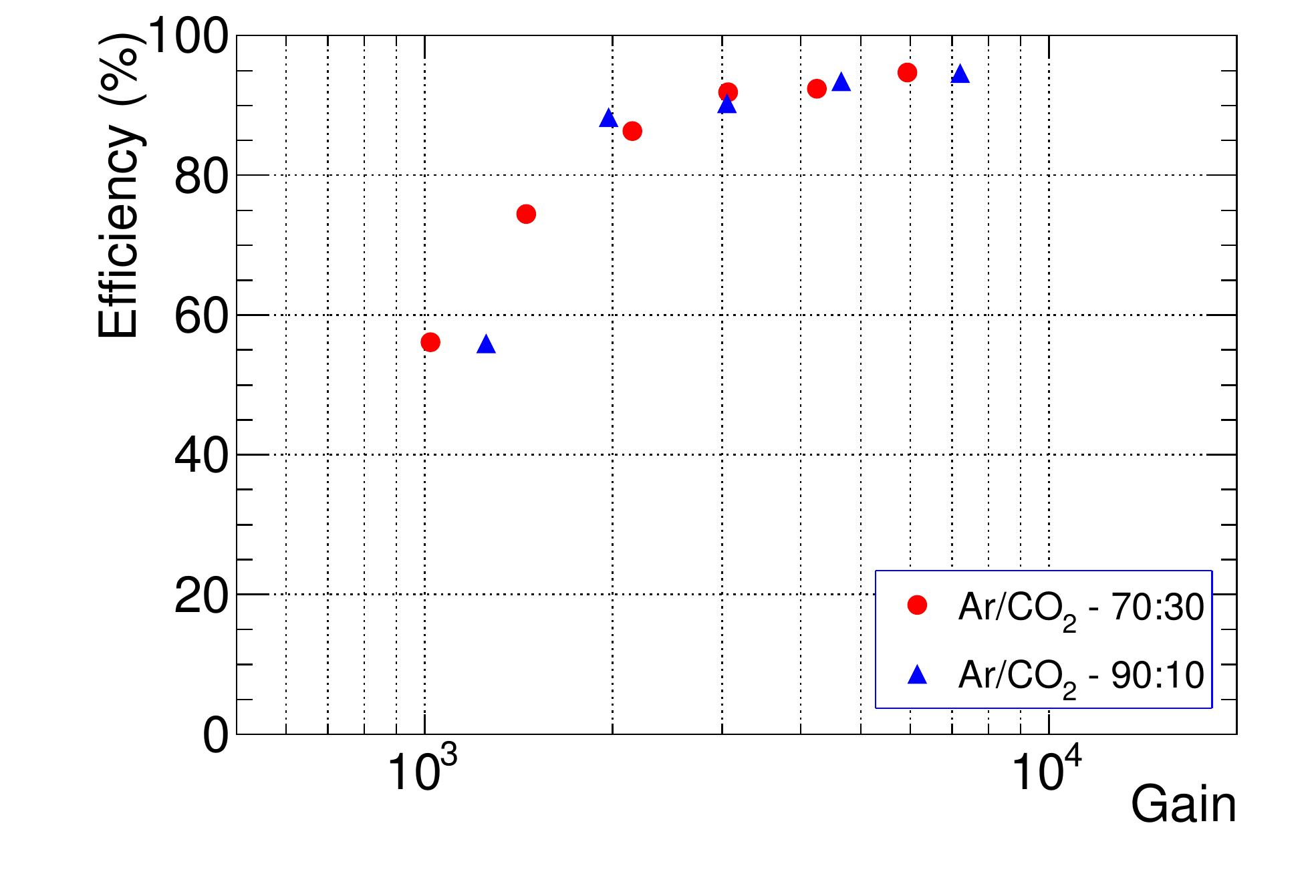}
\vspace{-5mm}
\caption{(a) Gain variation as a function of HV in Ar/CO$_{2}$ 70:30 and 90:10 gas, (b) Efficiency as a function of gain variation.}
\label{gain eff}
\vspace{-2mm}
\end{figure}

The gain of the detector is calculated from the mean ADC of the  $^{55}$Fe spectrum with a Gaussian fitting is shown in Fig.~\ref{gain eff}(a) as a function of HV. We find that in case of Ar/CO$_{2}$ 90:10 gas has higher gain at low operating voltage compared to the Ar/CO$_{2}$ 70:30 because of larger fraction of Argon content.

Efficiency of the detector should depend on the gain only providing all other setup condition same.  This is demonstrated in Fig.~\ref{gain eff}(b) where efficiency is plotted vs gain for the given HV values.
\vspace{-3mm}
\subsection{Time resolution}
\begin{figure}[!ht]
	\centering
	\includegraphics[width=2.35in]{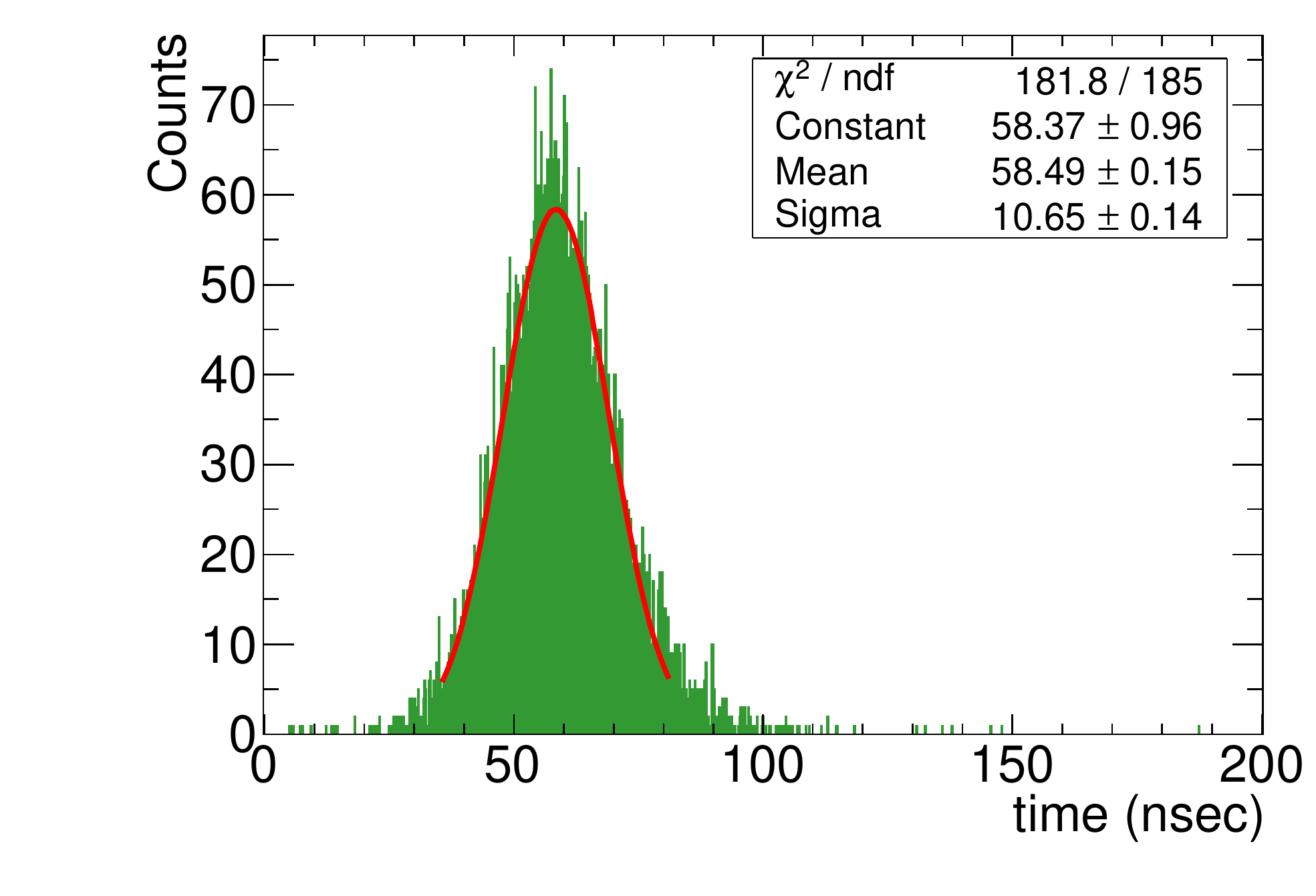}
	\includegraphics[width=2.35in]{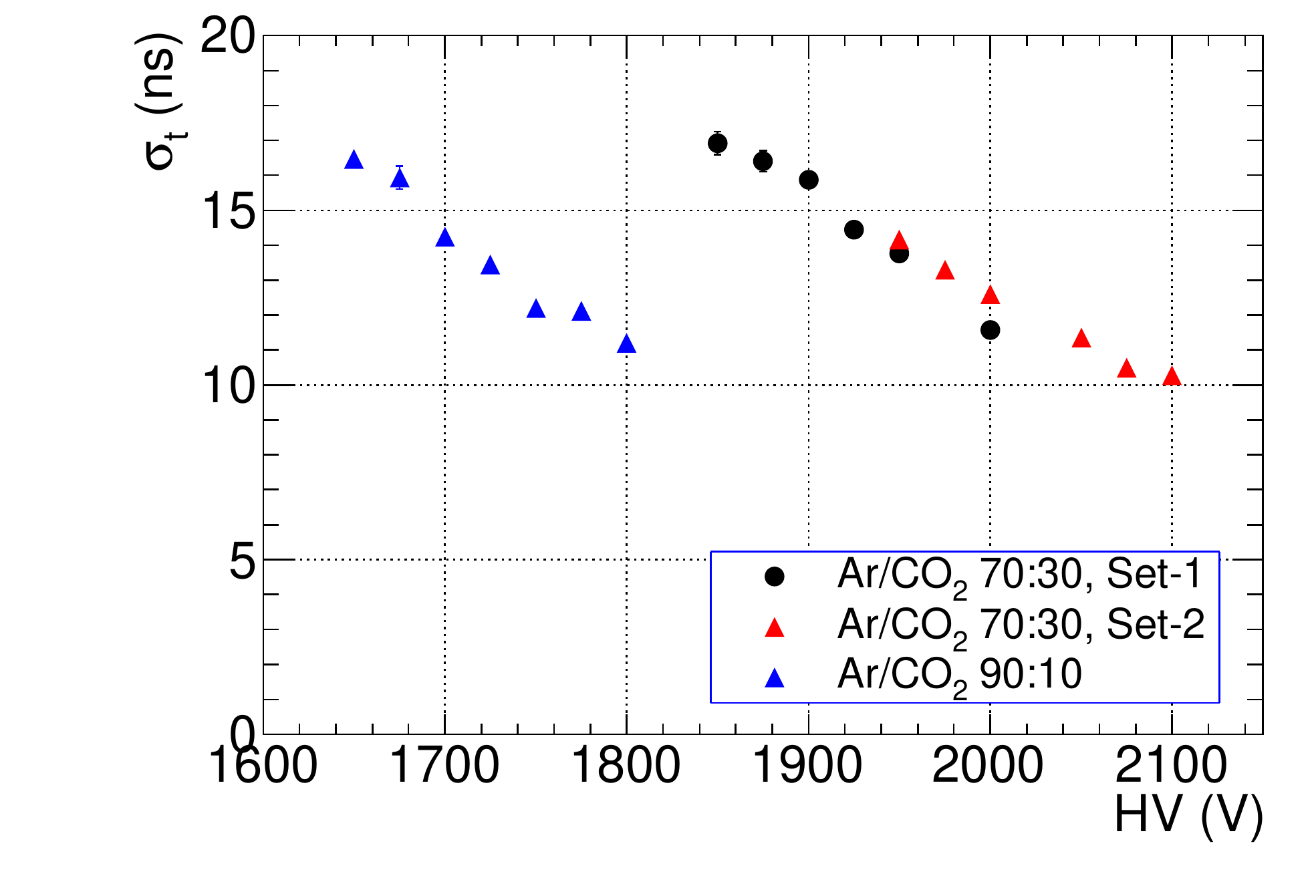}
	\vspace{-5mm}
	\caption{(a) Time spectrum at 2075~V in Ar/CO$_{2}$ 70:30 gas, (b) $\sigma_{t}$ variation as a function of HV.}
	\label{time reso}
	\vspace{-5mm}
\end{figure}

Time resolution of a detector is depends on spatial distribution of the primary charge cluster, electric field, gas and pressure. In this study time resolution is measured with both gas mixture.  This has been measured using the  3-fold scintillator signal as a start signal of a ORTEC TAC and the fast logic signal of the detector as stop signal. Fig.~\ref{time reso}(a) shows a typical time spectrum at 2075~V in Ar/CO$_{2}$ 70:30 gas. The time resolution ($\sigma_{t}$) as a function of HV is plotted in the Fig.~\ref{time reso}(b). For these gas mixtures we obtained $\sigma_{t}\sim$10ns.
\vspace{-6mm}
\section{Summary}
A number of small prototype MWPC detectors  has been fabricated and operated with Ar/CO$_{2}$ 70:30 and 90:10 gas mixtures. The efficiency, charge fraction, gain, timing resolution have been measured using radioactive sources. A comparative study of gain and efficiency has also been performed. Gas gain approaching 10$^4$ is achieved for both the gas mixtures. Time resolution of the detector is also measured with different gas mixtures at atmospheric pressure and $\sigma_{t}$ value of $\sim$10 ns is achieved.
\vspace{-3mm}
\section*{Acknowledgment}
RNP acknowledges the receipt of UGC-NET fellowship for this work and SB acknowledges the support of DST-SERB Ramanujan Fellowship. YPV thanks INSA for the Senior Scientist position.


\vspace{-3mm}
\begin{thebibliography}{99}
%
\bibitem {charpak1968}G. Charpak {\it et al.}, Nucl Instr. and Meth. {\bf 62}, (1968), 262.
%
\bibitem{chumedical}W. T. Chu {\it et al.}., Proc. SPIE {\bf 03772}, Physics and Engineering in Medical Imaging, (1982), 221; doi:10.1117/12.934519. 
%
\bibitem{lacymedical}Jeffrey L. Lacy {\it et al.}, J Nucl. Med. {\bf 25}, (1984) 1003-1012.
%
\bibitem{rajendra}R. N. Patra, {\it et al.}, Proc. DAE-BRNS Symp. on Nucl. Phys. {\bf 61}, (2016), 1052.
\end{thebibliography}
\end{document}